\documentclass{elsart}

\usepackage{natbib}

 \usepackage{graphicx}

\usepackage{amssymb}

\begin{document}

\begin{frontmatter}



\title{A systematic analysis of X-ray afterglows of gamma-ray burst observed by XMM-Newton}


\author[gendre]{Bruce Gendre}, 
\ead{gendre@rm.iasf.cnr.it}
\author[gendre]{Luigi Piro}, \&
\ead{piro@rm.iasf.cnr.it}
\author[gendre]{Massimiliano De Pasquale}
\ead{depasq@rm.iasf.cnr.it}
\address[gendre]{Istituto di Astrofisica Spaziale e Fisica Cosmica, via fosso del cavaliere 100, 00133, Roma, Italy.}

\begin{abstract}
This work is part of a systematic re-analysis program of all the data of Gamma-Ray Burst (GRB) X-ray afterglows observed so far, in order to constrain the GRB models. We present here a systematic analysis of those afterglows observed by XMM-Newton between January 2000 and March 2004. This dataset includes GRB 011211 and GRB 030329. We have obtained spectra, light curves and colors for these afterglows. In this paper we focus on the continuum spectral and temporal behavior. We compare these values with the theoretical ones expected from the fireball model. We derive constraints about the burst environment (absorption, density profile) and put constraints on their beaming angle.
\end{abstract}


\end{frontmatter}

\section{Introduction}
\label{intro}

While there is a growing sample of data for several tens of gamma-ray bursts (GRBs), we still know very little about their environment and progenitor. According to the standard model, GRBs are produced by a blast wave which propagates into the medium surrounding an unknown progenitor \citep{ree92, mes97, pan98}. The nature of this surrounding medium depends on the progenitor nature, and is supposed to be either an InterStellar Medium of constant density (ISM) or a wind profile arising from a massive star which decreases in density with the distance to the star (hereafter wind model, Chevalier \& Li 1999). Features such as ultra relativistic jets can complicate this model due to the symmetry change it can induce \citep{rho97}.

We have initiated a systematic analysis in a common way of all GRB X-ray afterglows observed so far in order to constrain the surrounding environment of GRBs. This can allow us to put constraints on the nature of the GRB progenitors. After the reduction and analysis of the Beppo-SAX observations \citep{pir04}, we present here the results we obtained with the XMM-Newton observations. This paper is organized as follow. We present the XMM-Newton observations and the data analysis in Section \ref{sample}. We expose our results in the Sections \ref{opt} and \ref{closure}. We discuss this results in the Section \ref{discu} before concluding.

\section{Sample of GRBs used and data analysis}
\label{sample}

   \begin{table}
      \caption[]{GRB X-ray afterglows observed with XMM-Newton. For each afterglow, we give the source name, the exposure duration and the net observation duration, together with the results of the spectral and temporal analysis. The X-ray absorption excess upper limits are given at the 90 \% confidence level.}
         \label{table1}
     $$
         \begin{tabular}{ccccccc}
            \hline
            \noalign{\smallskip}
Source  & Time   & Exposure & Net      & Temporal & Energy   & Excess of \\
name    & since  & duration & duration & decay    & spectral & absorption\\
        & burst  &  (ksec)  & (ksec)   &          & index    & ($10^{21}$ cm$^{-2}$\\
        & (ksec) &          &          &          &          &  at z=1)\\
            \noalign{\smallskip}
            \hline
            \noalign{\smallskip}
GRB 001025A& 162.5 &  40.8 &  33.4 & 1.2 $\pm$ 3.0   & 2.1  $\pm$ 0.5  & 6 $\pm$ 3\\
GRB 010220 &  53.9 &  44.5 &     0 &                 &                 &  \\
GRB 011211 &  43.5 &  33.6 &  33.6 & 2.1 $\pm$ 0.3   & 1.2  $\pm$ 0.1  & $<$1.4\\
GRB 020321 &  35.5 &  54.5 &  32.0 &                 &                 & \\
GRB 020322 &  53.7 &  29.0 &  28.4 & 1.3 $\pm$ 0.3   & 1.1  $\pm$ 0.1  & 6.5 $\pm$ 0.8\\
GRB 030227 &  29.8 &  10.9 &     0 &                 &                 &  \\
GRB 030329 &3200.4 & 244.9 & 143.5 & 2               & 1.0  $\pm$ 0.2  & $<$1.0 \\
GRB 031203 &  22.1 & 112.8 &  92.2 & 0.5 $\pm$ 0.1   & 0.8  $\pm$ 0.1  & 11 $\pm$ 9\\
GRB 040106 &  20.2 &  44.6 &  37.0 & 1.4 $\pm$ 0.1   & 0.49 $\pm$ 0.04 & $<$0.5 \\
GRB 040223 &  17.6 &  45.2 &  14.3 & 0.7 $\pm$ 0.3   & 1.7  $\pm$ 0.2  & 60 $\pm$ 16\\

            \noalign{\smallskip}
            \hline
         \end{tabular}
$$
NOTA : For GRB 030227, a process problem occurred with the SAS 6.0. We did not analyze it as we suspect the calibration to be false. GRB 020321 has no firm detection, we did not analyze it. The decay index of GRB 030329 is not constrained due to the very late observation time. The R magnitude 11 hours after the burst for those burst we studied (see text) are 23.25 (GRB 011211), 23.19 (GRB 020322) and 21.84 (GRB 040106). We also got an upper limit of 13.57 (GRB 040223), and one late observation with a significant supernovae contamination (R=20.14 @ 12 days, GRB 031203).

   \end{table}

We have retrieved from the archive the data of each GRB observed by XMM-Newton before the summer 2004. We have reprocessed and calibrated these data using the most up-to-date XMM SAS software (version 6.0) with the latest calibration files. We have excluded from these calibrated files any event occurring during a flaring background interval, using a very conservative condition (this can explain the difference between the exposure times reported in the literature and our analysis). We finally filtered the event files for good events with energy between 0.3 and 10.0 keV, and used the remaining events for spectral and temporal analysis. 

In this paper, we focus on the continuum properties. Hence, for spectral analysis we used the canonical power law model, letting the absorption vary as a free parameter. An analysis aimed to search for narrow features is ongoing. We rebinned the spectra in order to obtain at least 20 net counts in each bin. For the temporal analysis, we tried to fit the light curve with a simple power law model. Table \ref{table1} lists all our results.

\section{Comparison of the optical and X-ray fluxes}
\label{opt}

We compare the optical and X-ray fluxes of the XMM-Newton bursts together with a sample of burst observed with Beppo-SAX in Fig. \ref{Fig1}. All the values have been corrected for the galactic absorption (X-ray) or extinction (R band), using the work of \citet{sch98}. One can note that not all XMM-Newton bursts are displayed in this figure :  three GRBs in our sample do not have detected optical afterglows (or the detections are very late). We do not include the corresponding upper limits in Fig. \ref{Fig1} due to the poor constrains they put on them.

However, GRB 020322 is very interesting because it is located in the ``dark'' side of the figure : it is a normally bright X-ray afterglow, while it displays a faint optical afterglow. This burst also displays an excess of X-ray absorption \citep[see Table \ref{table1} and][]{wat02}. Using a galactic gas-to-dust law, \citet{wat02} indicated that this burst is optically extincted. Using the more correct dust-to-gas law of \citet{cal94} indicated in the work of \citet{str04},
 we calculated an extinction of 1.6 $\pm$ 0.4 in the R band. This value takes into account all the uncertainties of the spectral fit (i.e. on the galactic absorption, on the intrinsic absorption and on the spectral index). We present in Fig. \ref{Fig1} the best fit relationship obtained from the data, including (dashed line) and excluding (solid line) from the fit the upper limits. These relationships imply an extinction of $0.5-2.5$ for this burst, compatible with our finding from the X-ray.

\section{The closure relationships}
\label{closure}

 The spectral index and the temporal decay of a burst are linked together \citep{sar98, che99}. This relationship depend on the burst environment (wind or ISM), on jet features and on the cooling frequency position. This gives a set of six closure relationships which could give insights about the burst environment and/or jet features. We show in Fig. \ref{Fig2} the calculation result of the six closure relationships for the best constrained bursts from our sample. We can note that jet features are excluded most of the time. When we cannot exclude them, we also cannot exclude any other model. 
We can also note that due to a degeneration in two closure relationships, we cannot discriminate the burst environment from a wind profile or a constant ISM medium when the cooling frequency, $\nu_c$, is below the X-ray band. The only notable exception is GRB 040106, which should be, according to the closure relationship, surrounded by a wind profile. \citet{gen04} have shown that the optical observations agree with this interpretation and that the cooling frequency is above the X-ray band 6 hours after this burst.

\section{Discussion and conclusions}
\label{discu}

We have analyzed in a common way the GRB X-ray afterglows observed by XMM-Newton. Using spectral and temporal analysis, we have constrained the surrounding environment of some of these burst.

The first result is that the fireball expansion one day after the burst does not show jet features. Note that this does not rule out the possibility of a collimated fireball, but simply put a lower limit to the jet opening angle, which should be larger than $0.166 (n_1/E_{52})^{1/8}$ rad (where $n_1$ is the density in cm$^{-3}$ units and $E_{52}$ the total energy of the burst). This could imply a possible large beaming angle of gamma ray bursts.

Using the X-ray afterglow data only, we can constrain the burst environment in some case (e.g. GRB 040106, Gendre et al. 2004). But in most of the case, both wind profile and ISM medium can fit the data, and other observations (e.g. optical) are needed to refine the constraints.

\begin{figure}
   \centering
   \resizebox{8.5cm}{!}{\includegraphics{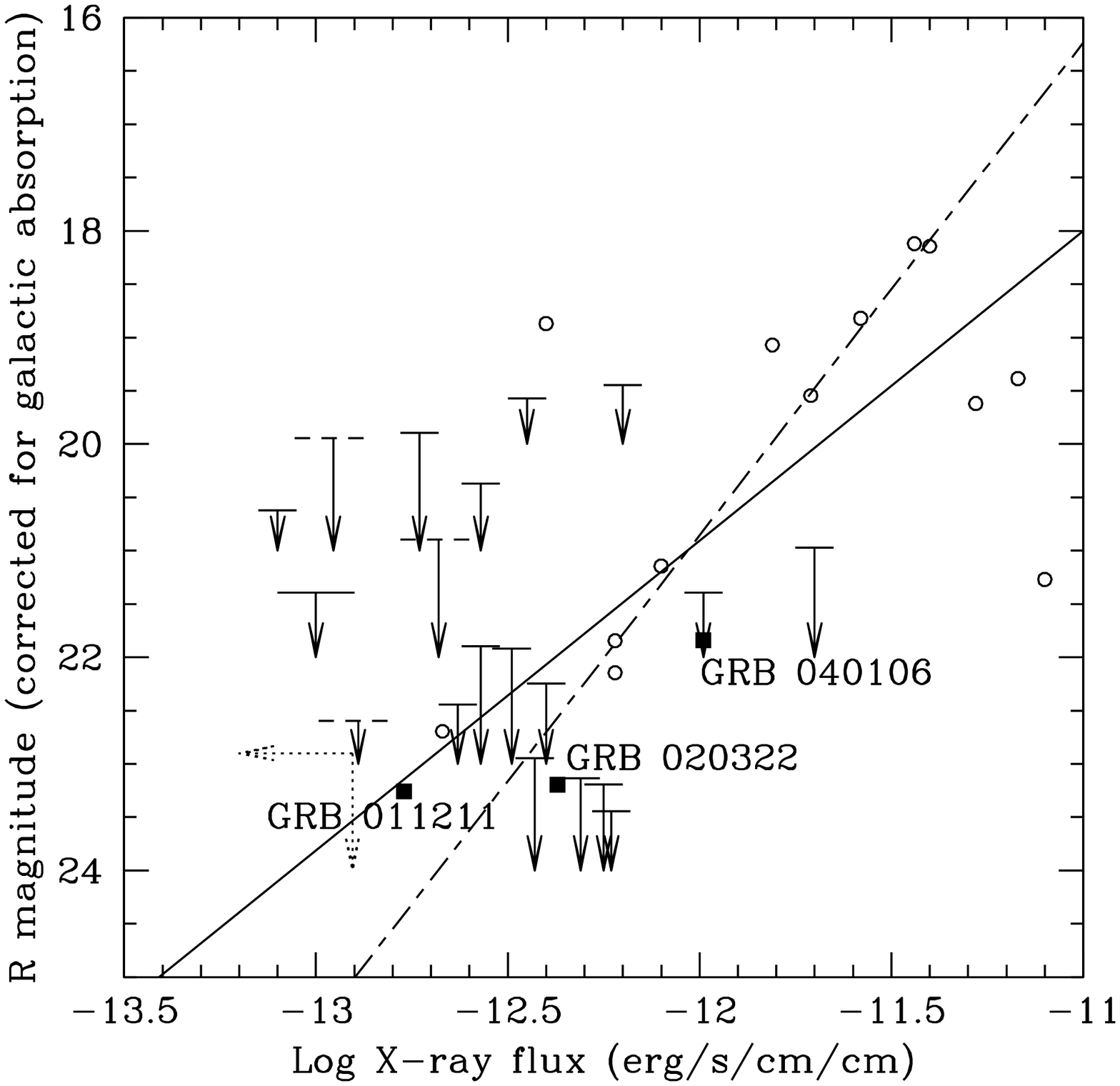}}
   \caption{Optical versus X-ray fluxes of GRB afterglows 11 hours after the burst. Squares and open circles represent XMM-Newton and Beppo-SAX data \citep[extracted from][]{pas03} respectively. Lines indicate the best fit relationships (see text for details).}
              \label{Fig1}%
    \end{figure}

\begin{figure}
   \centering
   \resizebox{10cm}{!}{\includegraphics{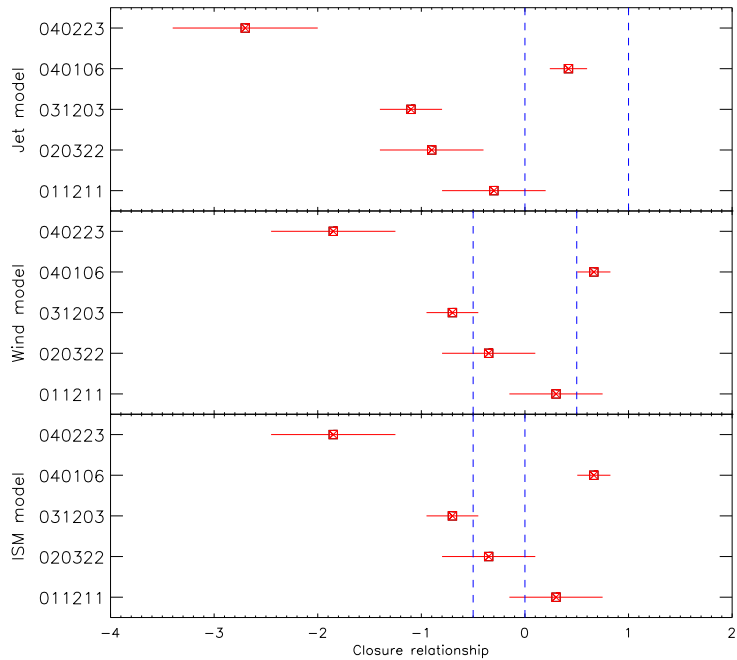}}
   \caption{Closure relationships of the XMM-Newton GRB afterglows. We have indicated the GRB name and the three model possibles (Jet, Wind and ISM). The vertical dotted lines indicate the theoretical values for each model.}
              \label{Fig2}%
    \end{figure}

Finally we have detected a significant excess of X-ray absorption in 3 bursts. We have shown that the GRB 020322 X-ray absorption implies an optical extinction which could explain its faint optical afterglow. Because this burst is located in the ``dark side'' of our Fig. \ref{Fig1} (where dark GRBs are located) it is very important for the study of dark GRBs : an X-ray absorption similar to the one measured in the case of GRB 040223 would have implied an optical extinction large enough to prevent the detection of any optical afterglow (making this burst dark).

Our systematic study of X-ray afterglows will allow us to complete the Beppo-SAX and XMM-Newton sample with the Chandra X-ray afterglows. Analyzed in a common way, these afterglows will give us the opportunity to better constrain the burst environment and thus to put constraints on the burst progenitors. The very next launch of the SWIFT experiment and its very fast localization abilities may also grow quickly the sample of X-ray afterglows observed with very sensitive experiments such as XMM-Newton, giving more insights about the burst environments and progenitors.




\clearpage

\end{document}